# Fractal Scaling of Cortical Matter, Amyloid Fragmentation and Plaque Formation across Rodents and Primates

J. C. Phillips

Dept. of Physics and Astronomy, Rutgers University, Piscataway, N. J., 08854

## Abstract

Thermodynamic tools are well suited to connecting evolution of protein functionalities to mutations of amino acid sequences, especially for neuronal network structures. These tools enable one to quantify changes in modular structure and correlate them with corresponding changes in observable properties. Here we quantify modular rodent-primate changes in amyloid precursor protein A4 and β amyloid fragments. These are related to changes in cortical connectivity and to the presence (absence) of plaque formation in primates (rodents). Two thermodynamic scales are used, descriptive of water/air protein unfolding (old), or fractal conformational restructuring (new). These describe complementary aspects of protein activity, at respectively higher and lower effective temperatures.

**Introduction**

Evolution optimizes protein functionalities with respect to environment. The physical and social environments of primates and rodents differ greatly, and these differences are partially represented by changes in amyloid amino acid (aa) sequences. One of the striking differences between rodents and primates is that the latter species have larger numbers of neurons. Among primates the human brain is scaled up in its cellular composition and metabolic cost, with a relatively enlarged cerebral cortex that is remarkable in its cognitive abilities and metabolism simply because of its extremely large number of neurons [1]. One can go further by scaling cerebral morphological properties, and one finds that rodent properties scale differently from primate properties [2].

In two earlier articles we discussed human amyloid properties and their connection with aa sequences [3,4] using a thermodynamic approach based on two hydropathicity scales. These scales are the standard water-air, obviously first-order, enthalpy-based KD hydropathicity scale [5] and the new fractal MZ hydropathicity scale, based on Bak self-organized criticality (SOC), which was derived from the differential geometry of solvent-exposed surface areas of > 5000 protein segments from the Protein Data Bank. [6]. Amyloid properties have been studied extensively, and the fragmentation of the amyloid 770 aa precursor protein A4 into 40,42 aa β amyloid is a nearly critical dynamic process. Here we use human-rodent A4 and βA differences to define more accurately the content of our thermodynamic methods [3,4]. These methods involve one-dimensional (1-d) hydropathic profiles, and are particularly simple for fibril-forming, amphiphilic βA.

**Results**

The overall human and mouse A4 profiles are shown in Fig. 1 here, with the sliding window width W = 21, which has worked well [3] in describing the cleavage of A4 at 672Asp. Only the MZ scale is used here and later, as it was shown in [3,4] to provide better resolution of A4 dynamics. Because of 97% sequence identity, the differences between the human and mouse A4 profiles are small, yet they contain systematic features. Notably the large ones near and below 50, and centered on 250 and 620, make the human profile more hydrophilic and occur near hydrophilic extrema. These make the human A4 more flexible and more biologically active.

The overall roughnesses or variances ʀ(W) for human and mouse are very similar, and the human function was discussed in [3]. There we identified a phase separation spinodal, with a spinodal tie line. The mouse tie line defined by its ʀ(W) may correspond to a slightly lower effective temperature. The > 3/4 hydrophilic acid region in humans 230EEEEVAEVEEEEADDDED248DEDGDEVEEEAEE260 undergoes only one mutation, D248V in mouse, but this produces the large difference shown in Fig. 2. The human W = 21 minimum at 107.5 is the second deepest – second most acidic - one known to us. The deepest is 94.8, found near the center of 413 aa Aspein, which is 3/5 Asp and < 0.1 hydrophobic, and is responsible for directed formation of calcite in the shell of the pearl oyster – an ultrasoft buffer supporting the growing ultrahard oyster shell [7].

Fig 3 shows the N-terminal region below 70. Here the human-mouse mutations are D64G and T15V. Human is more hydrophilic and more flexible near the two minima at 58 and 23. According to Uniprot, the latter is associated with the edge of the 1-17 signaling region, but it may be better to look on the N-terminal region as consisting of two hydrophilic hinges at 23 and 58, and two hydrophobic peaks (7LLLL10 and 39MFC41).

The principal focus of amyloid studies is the amyloid β segments βA = 672-712±1. The two human-mouse mutations R676G and H684R compensate each other in regions of MZ profiles where they overlap, as ψ(Gly) and ψ(His) are nearly equal on the MZ scale. There are two values of W to consider, W = 21 which describes fragmentation best [3], and W = 13, which best describes post-fragmentation folding of βA [1,4]. The differences between the human and mouse profiles for W = 21 and 13 are shown respectively in Figs. 4 and 5.

Looking at Fig. 4, we see that the overlap of the mouse mutations has stabilized mouse against fragmentation at 672. This is most satisfactory, as it is consistent with the absence of AD in rodents. In Fig. 5, we see another interesting point, because the stability of the sandwich folding should depend on the difference between the hydrophobic maximum at the 712±1 edge of the $\beta_1$ strand and the hydrophilic minimum at 673. This difference is reduced by 40% in mouse. It should also affect fibrillar formation, as well as the concentrations of amphiphilic Aβ [8].

The mouse mutation 670KM671 to 670NL671 gave rise to AD in mice which also overexpressed the isoform 695 [9]. This remarkable correlation involves large-scale interactions with local mutations. It raises the question of the mechanisms which stabilize the other major amyloid isoforms 695 and 751, relative to canonical 770. These mechanisms become clear if we analyze the reconstructions involved in the 695 and 751 isoforms. The 695 isoform differs from 770 by the deletion of 290-364. We therefore study the connection across 289 to 365 to understand why this human isoform is marginally stable, and gives AD in mice.

Our idea is that we can infer the best tuned value of W by smoothing the water wave density at the junction, that is matching ψ(289, W) to ψ(365,W). Fig. 6 shows the results with both the MZ and KD scales. Note the secondary dip near W = 9. This is probably related to matching the alpha helicities. However, with the MZ scale, W = 21 (membrane length scale) wins out, with the smallest difference and best match. Note that the multiple crossings of the KD profile are unacceptable, as the resulting 695 would be unstable at multiple length scales.

The water wave density matching condition is the same as the one that has been developed in dynamical applications of Voronoi partitioning and has been called "level set theory" [10]. The 75 aa deletion is large: how level are these boundaries? Fig. 7 shows a comparison of the matching with W = 21 compared to W = 41, using the MZ scale; it dramatizes the success of the W = 21 matching, and it also shows how large are the profile differences between the two values of W. Given how large the 74 aa deletion is, one might have expected W = 41 to give a good match, but W = 21 (membrane length) is much better.

The 751 isoform involves a much smaller deletion, 346-364 from 770, and it involves the same end point, 365. Which value of W is best here? According to Fig. 8, the W = 7 helical instability dominates here, and possibly there is even a helix formed across the 345-365 weld. What happens to cause the sign reversal in Fig. 8 at W = 7? Fig. 9 compares the W = 5,7,9 MZ profiles. The flat hydrophilic region 345-365 is backed up against a very strong 340 hydrophobic peak – the 290 end of the 75 aa deletion is far off to the other side of this peak. The deleted hydrophilic region 346-364 appears to be disordered, compared to the stable 340 peak.

Before we leave the subject of the central 345-365 weld, it is amusing to compare the human A4 profile with the monkey A4 (P53601) profile. The two sequences are 99% identical, and the mutations occur only in the 345-365 weld range. The differences in MZ9 weld range profiles are shown in Fig. 10. The human profile is smoother, and the deep hydrophilic hinge near 355 is strengthened in humans compares to monkeys, which appears to be evolutionarily advantageous for the larger and more complex human neuronal network. An additional advantage of the human A4 is the improved leveling or balancing of the mutated 355 hinge against the conserved 330 hinge [10], as well as the corresponding leveling of the 340 hydrophobic peak against the conserved 320 hydrophobic peak. It is quite remarkable that these extremely small differences between human and monkey A4 manifest themselves so exactly in both sets of extrema of the W = 9 MZ profile. Repeating the procedure used to derive Fig. 10 with the KD (first-order) scale, we find W = 7 is optimal. However, even this choice causes tilts of adjacent extrema ~ 20 times larger than those seen in Fig. 10 for the nearly level MZ pairs. If there are any "rounding errors" in Fig. 10, they are extremely small.

We can now discuss the remarkable emergence of AD in mice that combine the paired mutation 670KM671 to 670NL671with overexpressed isoform 695 [9]. The effects of the paired mutation alone are small (Fig.10), and the overexpressed isoform alone does not cause AD in mice; only the combination does. Fig. 11 shows that the paired mutation enhances hydrophobic interactions near 672, while Fig. 1 shows that the deleted segment 290-364 of 770 is a hydroneutral peak separating two deep hydrophilic minima. Removal of this peak leaves an unstable region that could bind to the hydrophobically enhanced paired mutation 670NL671, and facilitate Aβ fracture.

Comment: here we have focused on human-mouse A4 profile differences. A rat A4 sequence, P08592, is also available, which exhibits small profile differences with the mouse sequence P12023. If the degu A4 sequence were known, these differences would be quite interesting [11].

.

**Discussion**

Perhaps the first point to be made here is that our detailed analysis of mouse-human amyloid similarities and differences has revealed two characteristic values of W, W = 21 and W = 7. These are the same values as we identified in [3] in terms of a van der Waals equation of state, with $W = W_c = 21 = 3V_m$ , with $V_m = 7$. Thus the 695 (W = 21) and 751 (W = 7) isoform deletions of A4 770 are seen as examples of the general vdW model and level set principle [10]. The water waves we discuss exhibit membrane-scale (W = 21) features repeatedly (all A4 figures), as one would expect for a cell surface receptor [12].

Here we have continued the discussion of amyloid precursor protein A4 and amyloid plaque former Aβ using hydropathic scaling, which was begun in [3,4] for humans, to examine human-mice differences connected with the absence of AD in mice, and its partial emergence in transgenic mice. As before, we see systematic scaling trends which have both thermodynamic and mechanical or elastic interpretations. How far do these scaling ideas extend? Is it possible that these molecular ideas even extend to macroscopic dimensions, where rodent and primate cortical scaling are different [1,2]?

Indeed it is. The picture we have so far is that human molecular elements are more flexible than mouse molecular elements, especially near the two major hinge regions (249 and below 70) (Fig. 2 and Fig. 3), where the human hydrophilic valleys are deeper and narrower. This implies larger curvatures and more precise connectivity patterns, which are consistent with enhanced cortical folding in humans compared to rodents [2]. Human dendritic patterns also exhibit self-similarity, which can be explained by maximizing the repertoire of possible connectivity patterns between dendrites and surrounding axons while keeping the cost of dendrites low [13]. Loss of functional connectivity is characteristic of AD [14]. The very strongly acidic region 230-260 is also probably disordered, and it is this disordered and strongly acidic flexibility that facilitates optimization of connectivity through modularity [15]. Such optimization is close to biomineralization [16].

**Conclusion**

Our analysis has revealed that the 97% sequence similarity of human and mouse 770 aa A4 contains a few mutations that critically affect cerebral functions and exposure to plaque formation. Inasmuch as the backbones of human and chicken 139 aa lysozyme *c* are indistinguishable at 1.5 A resolution, even though there is only 50% sequence similarity [17], one can confidently conclude that hydropathic analysis using the Brazilian fractal SOC scale [6] has brought us insights into balanced neural networks unobtainable by other methods.

In a Science Letter celebrating the journal's 150$^{th}$ anniversary, Mandelbrot stressed that fractals are not everywhere, and in particular that linear log-log fits over half a decade or less may be of little value [18]. The new feature of the MZ hydropathic scale [6] is the linearity against logN of

logϕ(aa,N) for all 20 aa for $4 \leq N \leq 17$, that is, over 0.6+ decades., Here ϕ(aa,N) is the solvent-accessible (to a 2A° water sphere) surface area ϕ(aa) of (2N + 1) aa segments. By itself this linearity for a single aa is not impressive, and it could be accidental. However, the same linearity is obtained, over the same range, for all M = 20 aa (but with different slopes, denoted by ψ(aa)).

One can examine the sensitivity of key features of protein profiles to modifications of the MZ scale. So long as these preserve the hierarchy of the MZ scale, and do not involve interchanging two ψ(aa), the modifications maintain the correlations between the original scale and the modified scale to > 0.995. In that case all the leading results quoted here and elsewhere are preserved. Hierarchically conserved ordering gives a lower bound for the equivalent single fractal range from information theory. It suggests that the number of equivalent decades spanned is at least log(17/4)(MlnM –M), with M = 20, or ~ 25 decades, surely a record. Perhaps everything in nature is not fractal, but many aspects of protein globular functions are.

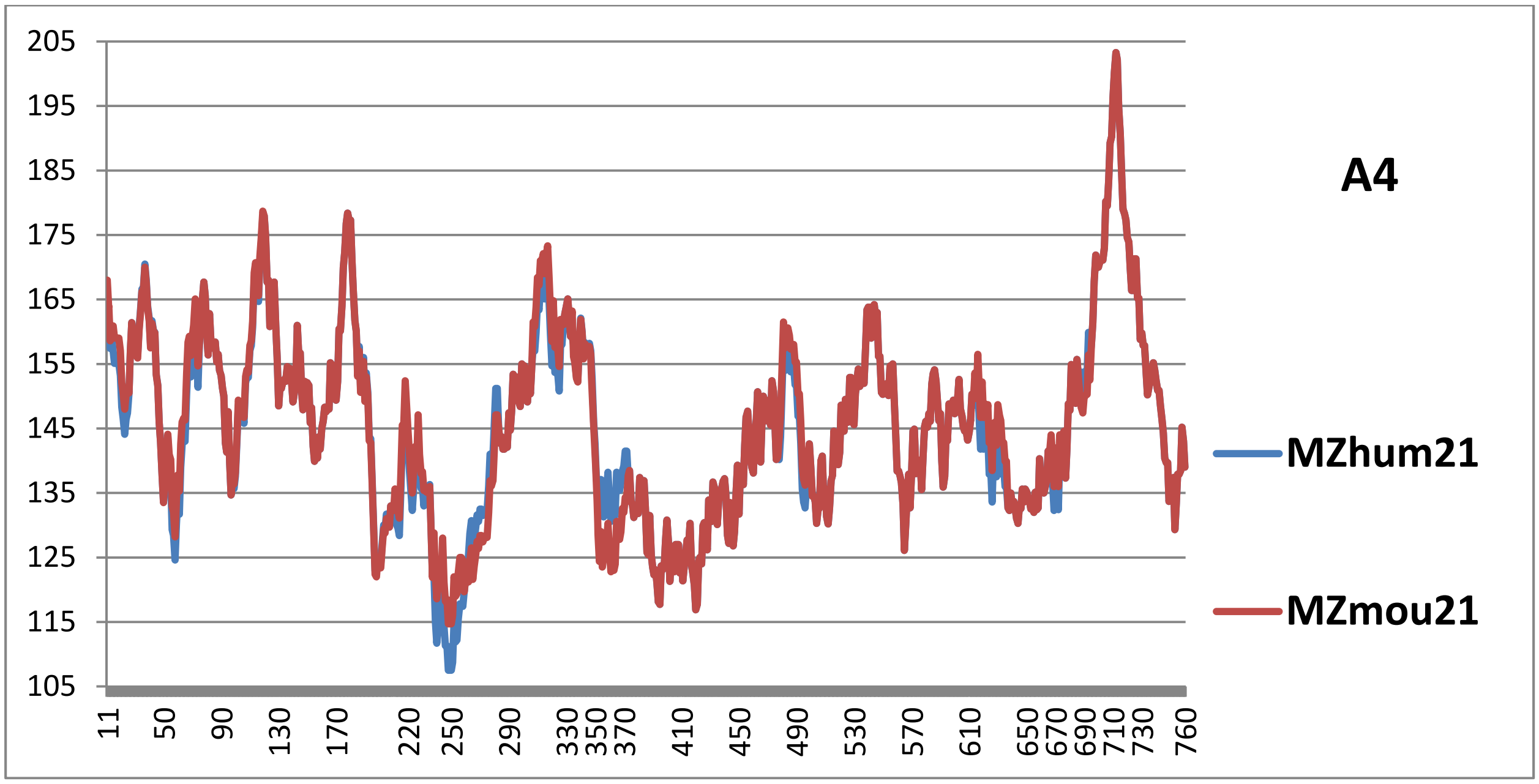

Fig. 1. Large-scale hydropathic profiles of ψ(aa,21) of the amyloid protein A4 for humans, P05067, and mice, P12023, using the MZ scale on the membrane modular length scale W = 21. The differences appear small, but are functionally significant. Most of the differences, especially the large ones near and below 50, centered on 249 and 620, make the human profile more hydrophilic and occur near hydrophilic extrema, thus increasing human A4 flexibility and biological activity. The secondary minimum near 350 is stabilized in humans by 6 partially compensating mutations.

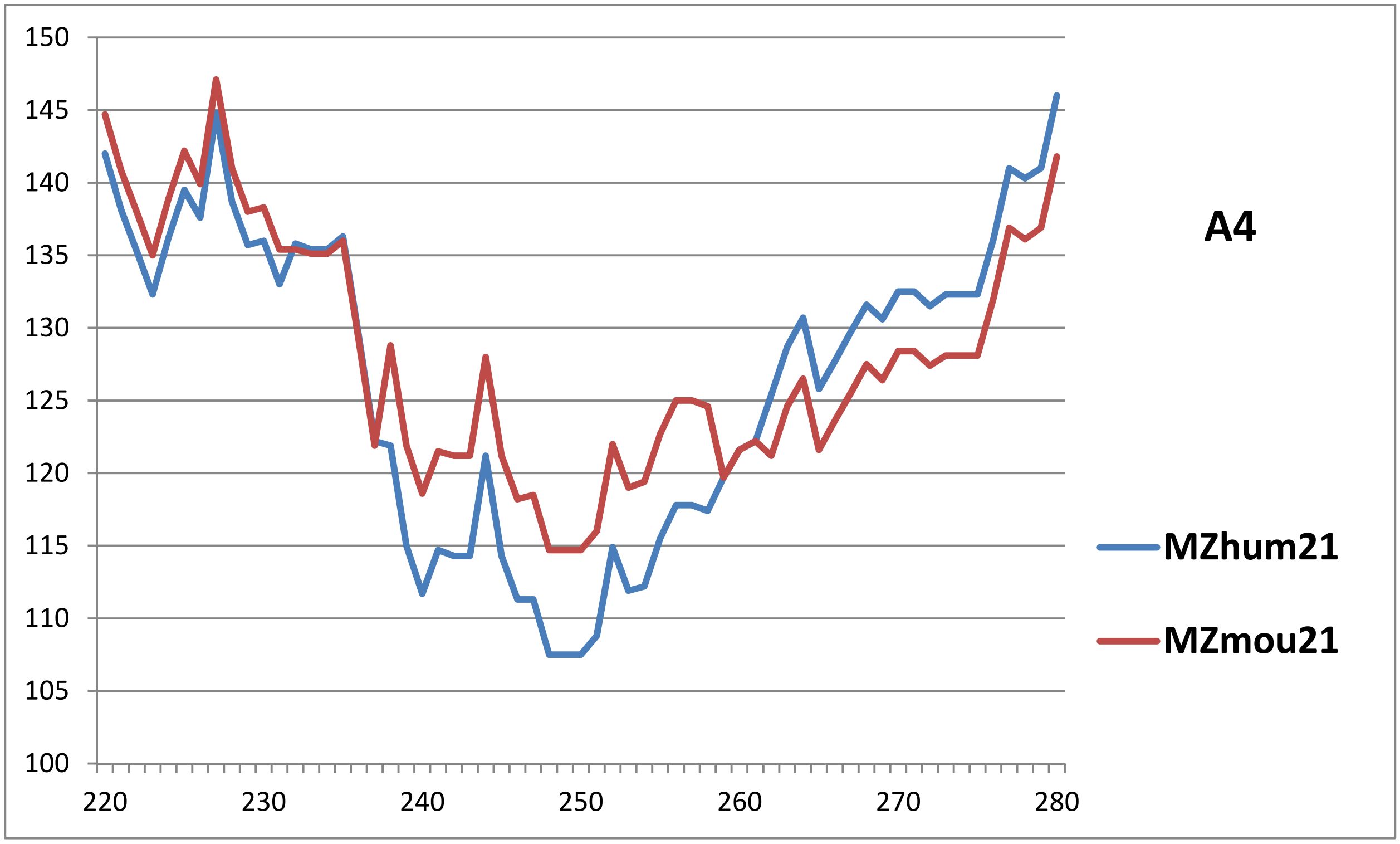

Fig.2. The deepest minimum of ψ(aa,21) is a hydrophilic hinge centered on 249 which is deeper and narrower in humans than in mice. A single mutation HumNMou, D248V, stabilizes A4 in mouse, while the human hinge is narrowed by the I272T mutation, which stabilizes the more extensive human network. This minimum is associated with the segment 230-260, which contains 24 Glu and Asp and thus is richly hydrophilic and extremely flexible.

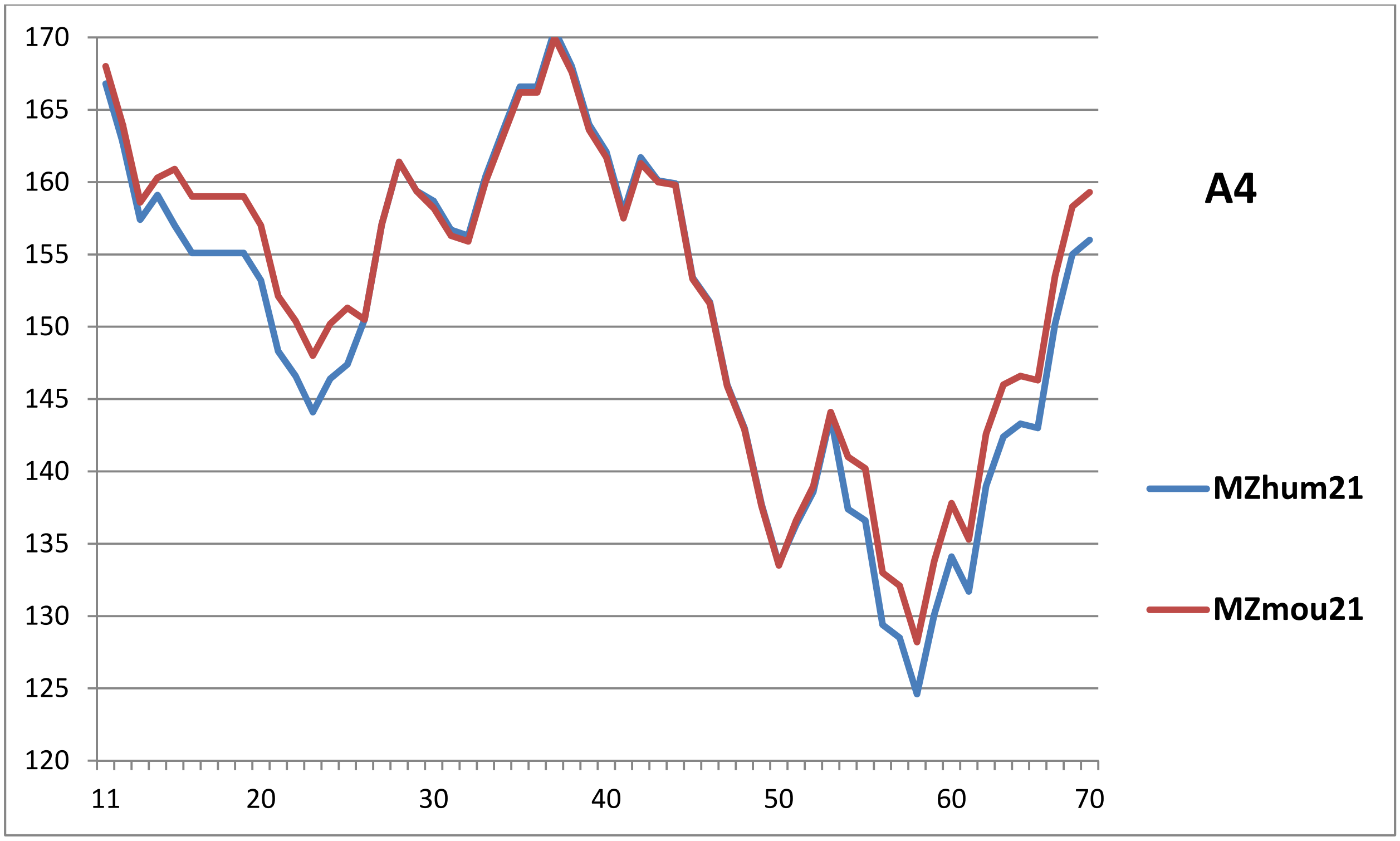

Fig. 3. Differences in N-terminal region (sites below 70) of ψ(aa,21) for human and mouse. The differences between human and mouse are concentrated near the 20 and 60 hinges. The two human hinges are more flexible than the mouse hinges.

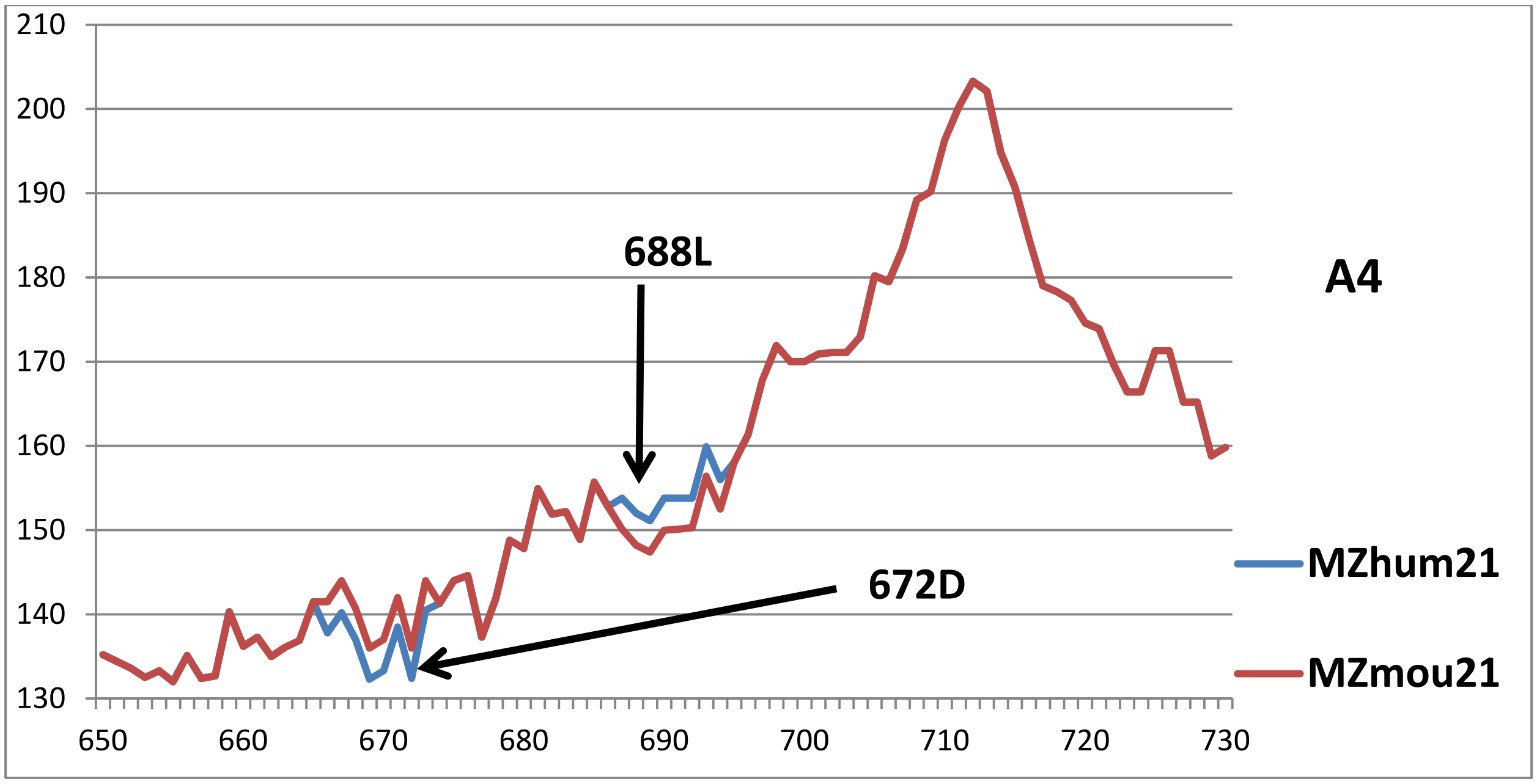

Fig. 4. Because of the close spacing of the human-mouse mutations at 678 and 684, their ψ(aa,21) profile differences are shifted to centers at 672 and 688.

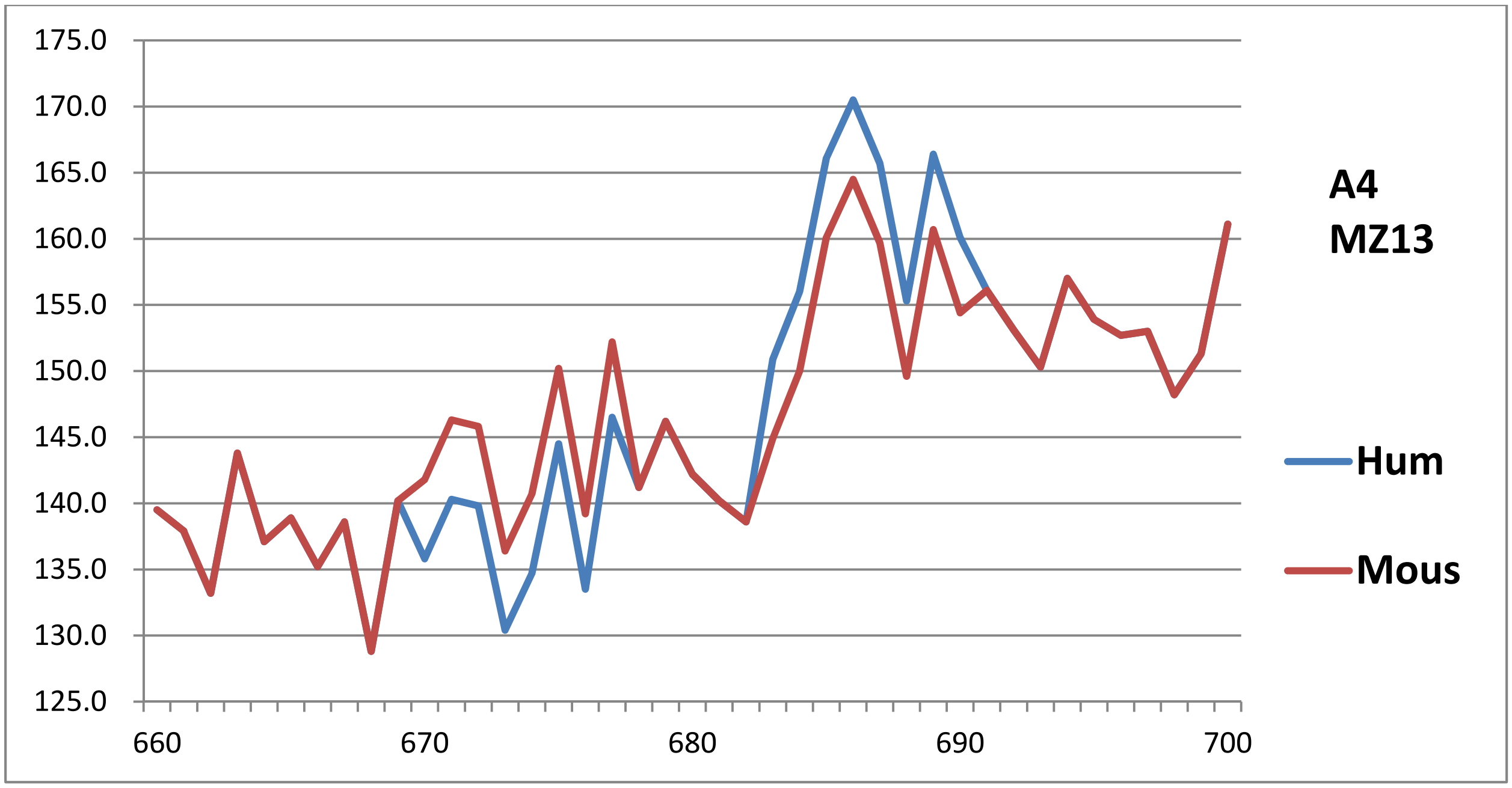

Fig. 5. On the smaller scale of W = 13, which better describes Aβ folded into the sandwich structure [1,4], the human minimum at 673 is much weaker in mouse, as is the hydrophobic edge of the sandwich at 688L. The difference between these two extrema is 40% smaller in the mouse profile.

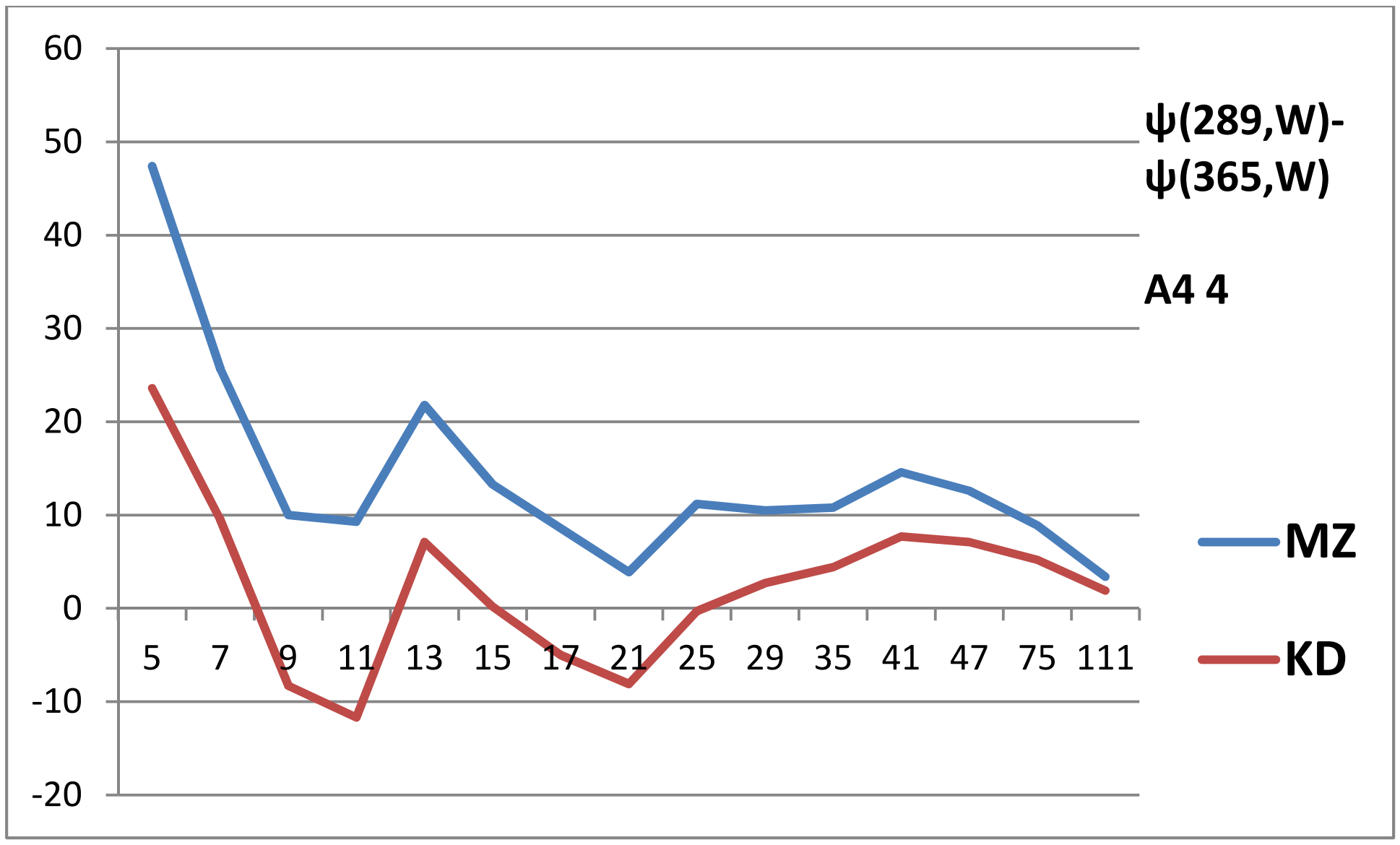

Fig. 6. The mismatch across the 75 aa 770 deletion 290-364 that occurs in isoform 695.

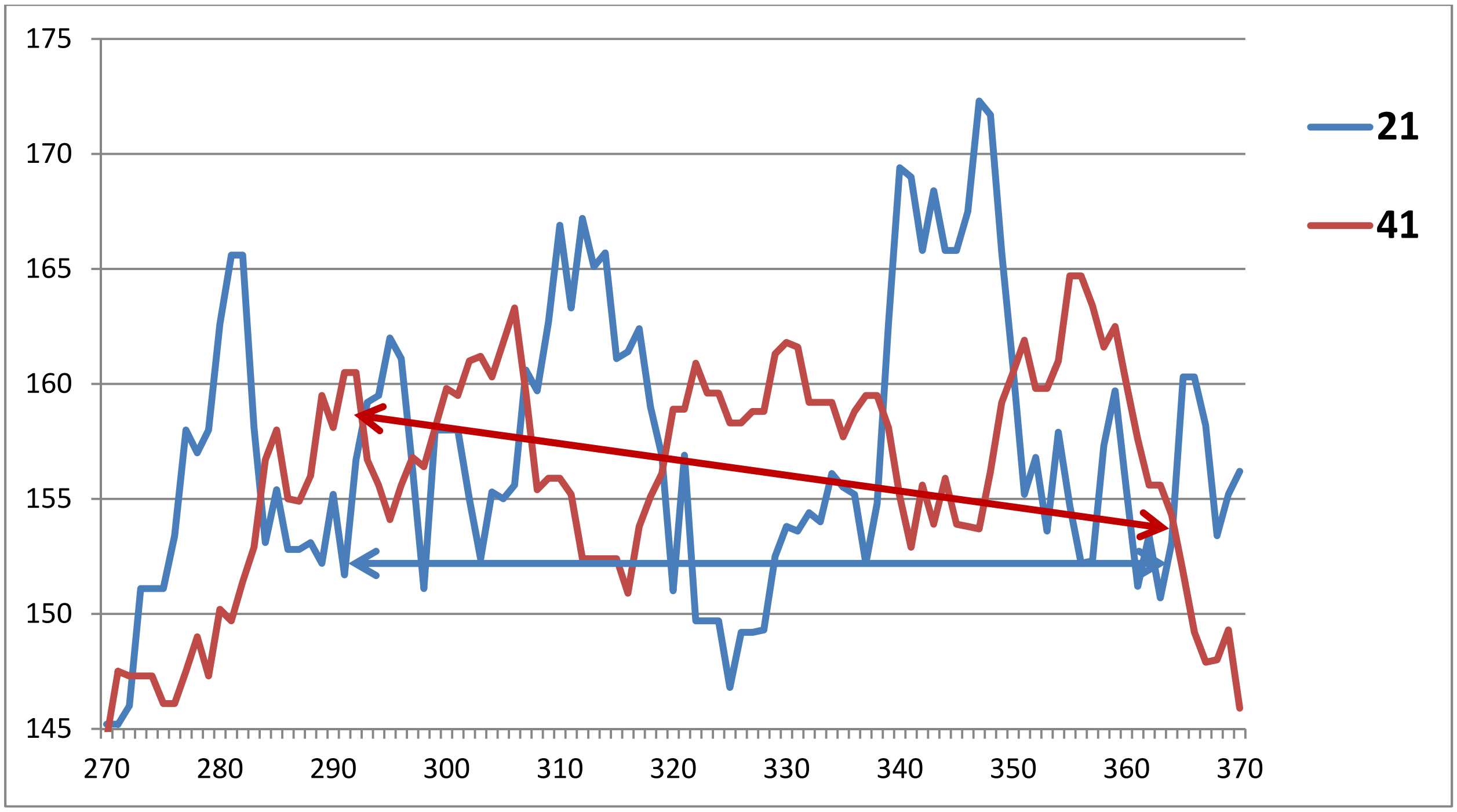

Fig. 7. "Level set" matching across the Isoform 695 deletion 295-364 for two values of W, 21 and 41, with the MZ scale (see Fig. 6 also).

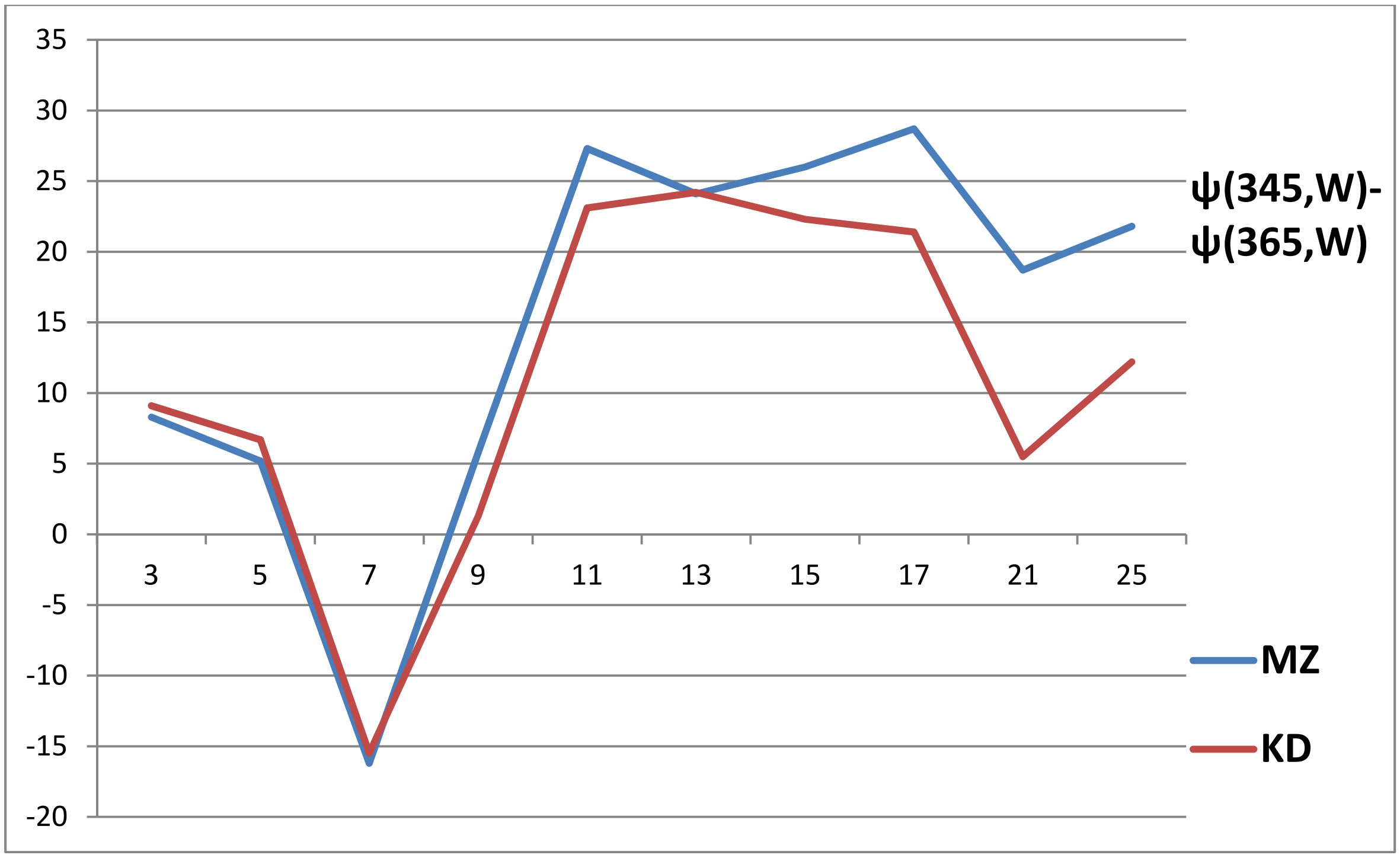

Fig. 8 Matching endpoints across the Isoform 751 deletion shows a clear-cut instability associated with W = 7 for both MZ and KD scales.

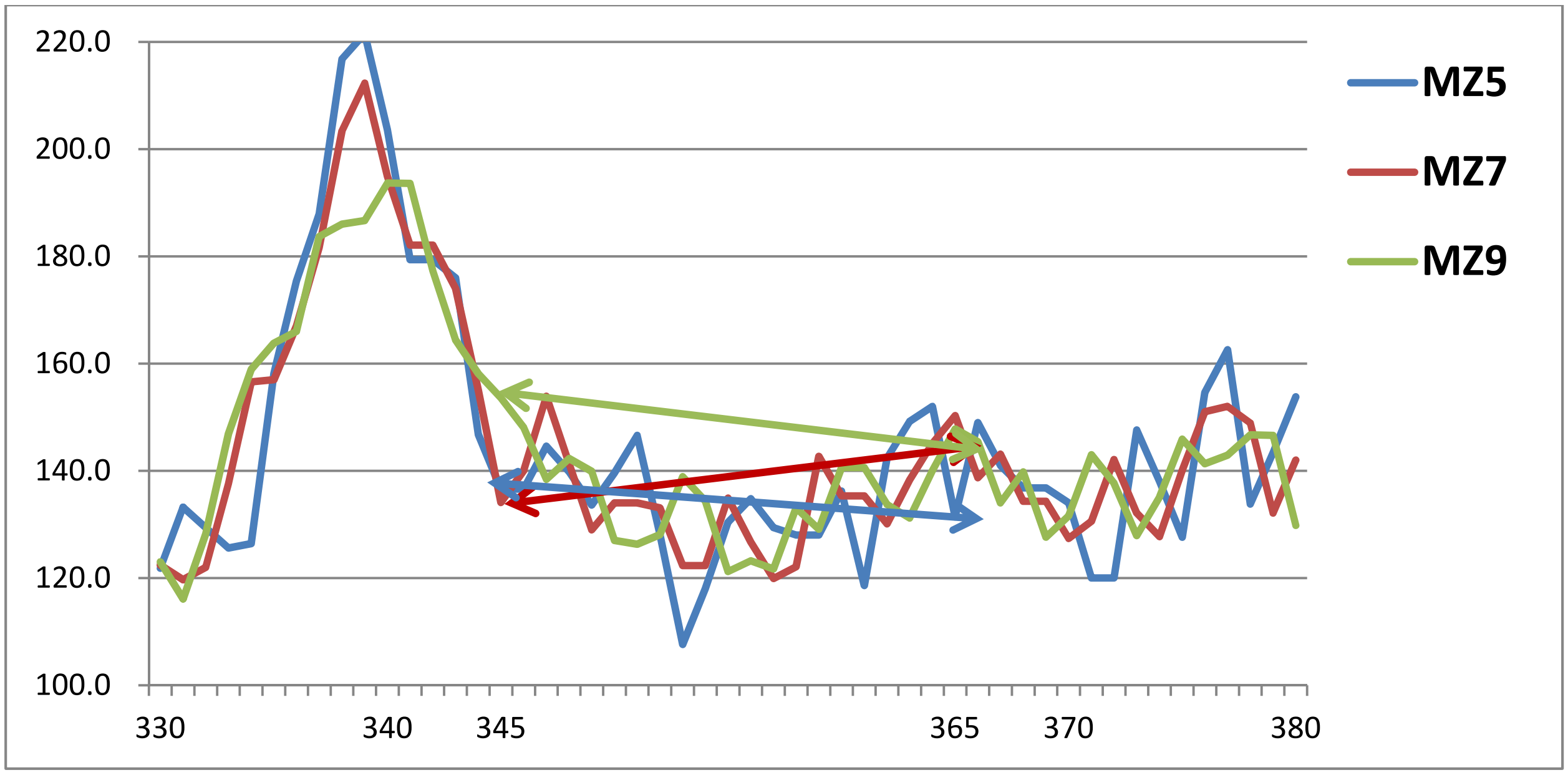

Fig. 9. The ψ(aa,W) profiles with the MZ scale are shown for three values of W. All three are nearly level, but the sign of the slope of the matching arrow for W = 7 is reversed from that for W = 5 and 9. See also Fig. 8.

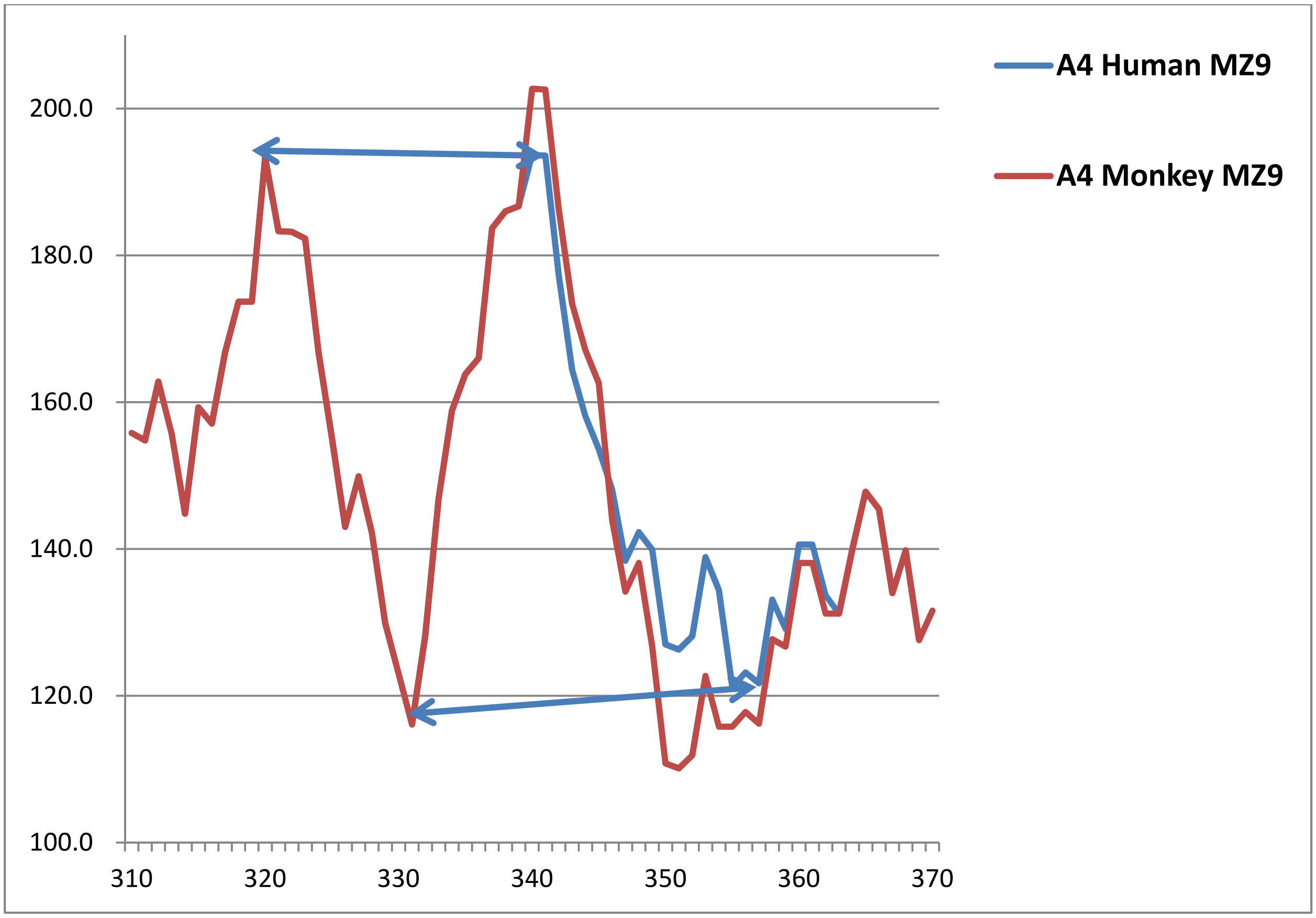

Fig. 10. Human and monkey A4 mutations are confined to the range 345-365 of the Isoform 751 deletion. In Fig. 8 we saw that W = 7 gives the best matching of MZ and KD endpoints across the 345-365 Isoform weld. For MZ monkey and human variances R(W), the largest difference occurs for W = 9, so this value of W has been used to profile A4 in this mutational range.

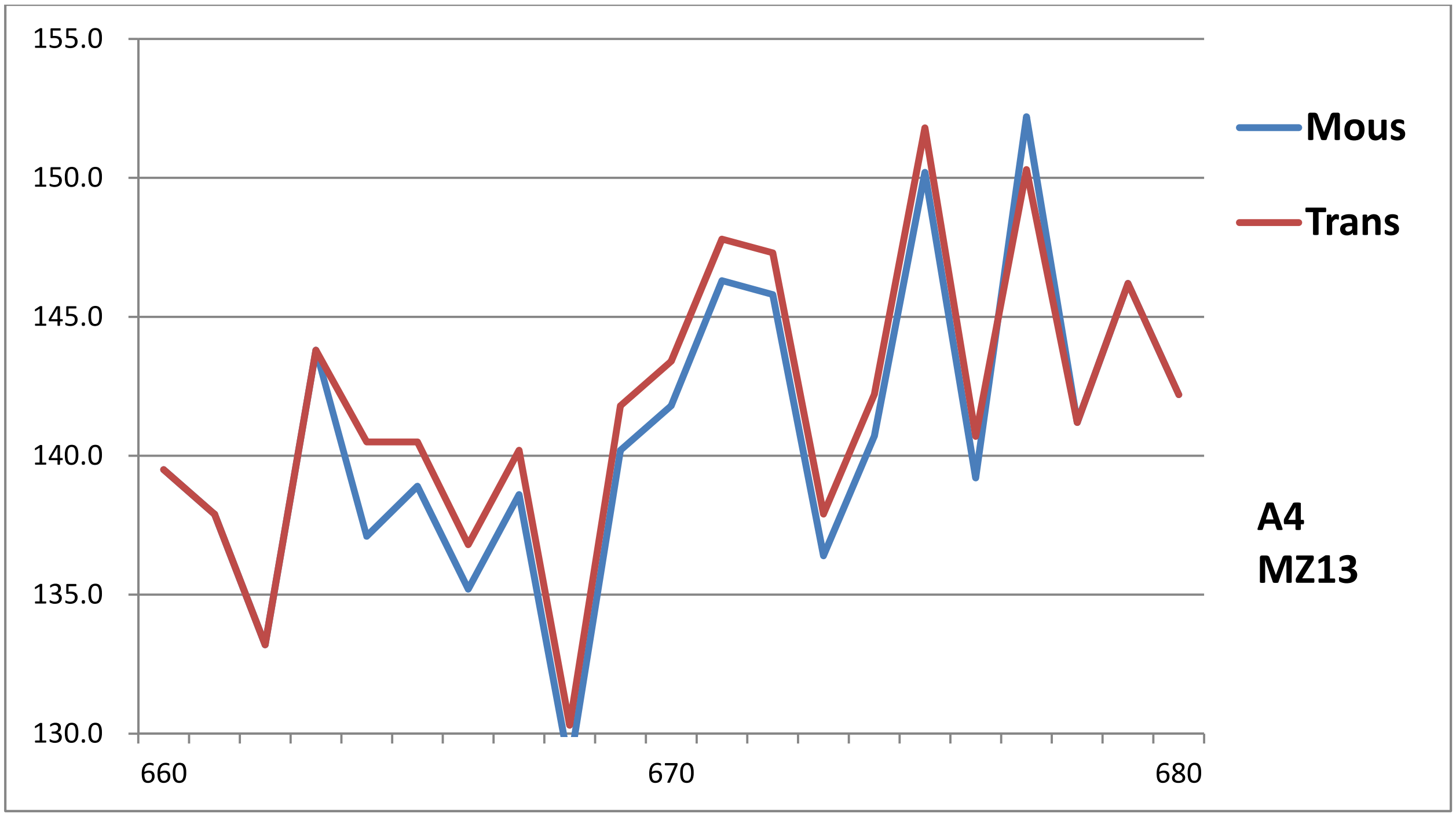

Fig. 11. The transgenic mutation 670KM671 gives rise to AD in mice which also overexpressed the isoform 695 [9].